\documentclass[prl,twocolumn,superscriptaddress]{revtex4-1}
\usepackage{epsfig,amsmath,amsfonts,amsthm,graphicx}
\usepackage{ulem}
\usepackage{color}
\newcommand{\st}[1]{#1^{st}}

\begin{document}

\title{Low-dimensional description for ensembles of identical phase oscillators subject to Cauchy noise}
\author{Ralf T\"onjes}
\affiliation{Institute of Physics and Astronomy, Potsdam University, 14476 Potsdam-Golm, Germany}
\author{Arkady Pikovsky}
\affiliation{Institute of Physics and Astronomy, Potsdam University, 14476 Potsdam-Golm, Germany}
\affiliation{Department of Control Theory, Nizhny Novgorod State University, Gagarin Avenue 23, 603950 Nizhny Novgorod, Russia}

\begin{abstract}
We study an ensembles of globally coupled or forced
identical phase oscillators subject to independent
white Cauchy noise. We demonstrate, that if the oscillators are forced in several harmonics, stationary synchronous regimes can be exactly described with a finite number of complex order parameters.
The corresponding distribution of phases is a product of wrapped Cauchy distributions.
For sinusoidal forcing, the Ott-Antonsen low-dimensional reduction is recovered.
\end{abstract}
\maketitle
%
%
%

One of the central challenges in synchronization theory is finding a possibility to describe the dynamics of populations of globally
coupled oscillators in terms of a few order parameters. Indeed, generally, following the 
pioneering approach of Kuramoto~\cite{Kuramoto-84},
one derives self-consistency conditions 
for the order parameters, which are formulated as integral equations and constitute an infinite hierarchy of coupled nonlinear relations. 
Such  self-consistency equations have also been derived for oscillators with coupling heterogeneity~\cite{Pazo-Montbrio-11}, or for identical oscillators subject to Gaussian white noise and coupling in the first or higher harmonics~\cite{Daido-96,Vlasov-Komarov-Pikovsky-15}. Self consistency means that given a stationary prior phase distribution, one calculates the force acting on oscillators of frequency $\omega$, the phase distribution of these oscillators and, by averaging over a frequency distribution $g(\omega)$, the phase distribution of the whole oscillator ensemble. The latter must be equal to the prior distribution. The self-consistency condition is often formulated in terms of integrals which have to be evaluated numerically.

Only in the case of coupling in the first harmonics, Lorentzian frequency distribution and zero noise strength the integrals can be evaluated explicitly. A Lorentzian frequency distribution is also central to the low dimensional dynamics of oscillator ensembles on the Ott-Antonsen manifold~\cite{Ott08}.
Indeed, this is the main reason a Lorentzian frequency distribution is used by default in many studies on nonidentical phase oscillators. 

Although Gaussian white noise acts qualitatively similar to frequency heterogeneity, for ensembles driven by Gaussian noise there is no exact low-dimensional reduction, except for several approximate approaches based on moment closures of the infinite hierarchy~\cite{Sonnenschein-Schimansky-Geier-13,Hannay-18,Goldobin_etal-18}.
As we demonstrate in this letter, the situation is different  when the noise is not Gaussian but L\'evy stable with exponent $\alpha=1$, i.e. Cauchy noise. Using Cauchy white noise instead of Gaussian one simplifies the analysis in a similar way as using a Lorentzian instead of a Gaussian frequency distribution, while keeping the bifurcation scenario qualitatively similar. In the simplest case of purely harmonic coupling, Cauchy noise allows for the Ott-Antonsen reduction (cf.~\cite{Tanaka-20}). Furthermore, we demonstrate that a low-dimensional reduction is also possible for \textit{multi-mode coupling}, albeit being restricted to stationary distributions which we show to be fully characterized by a finite number of modes.

We start with formulating general equations for an ensemble of identical phase oscillators driven by independent noise forces
\begin{equation}\label{Eq:PhaseEquations}
	\dot\vartheta_n = F(\vartheta_n) + \eta_n(t).
\end{equation}
For ensembles of coupled oscillators the driving force $F(\vartheta)$ depends on the phase distribution 
$P(\vartheta,t)$, which makes the problem of 
finding a stationary phase distribution nonlinear.
This is for instance the case with Kuramoto-Daido mean field coupling
\begin{equation}\label{Eq:KuraDaidoMF}
    F(\vartheta) = \int_0^{2\pi} H(\vartheta-\vartheta')P(\vartheta')d\vartheta' .
\end{equation}
The noise terms $\eta_n(t)$ are assumed to be Poisson processes of delta pulses with rate $\nu$ and amplitude distribution $W(\Delta\vartheta)$. 
Then the evolution equation for the phase density is given by the integro-differential equation
\begin{equation}\label{Eq:Continuity}
	\partial_t P 	= -\partial_\vartheta \left(F(\vartheta) P\right)+ \nu \int_0^{2\pi} \left(W(\phi)-\delta(\phi)\right)P(\vartheta-\phi)d\phi.
\end{equation}
As special cases we will consider a wrapped Gaussian distribution of pulse amplitudes $W^G\sim N(0,2D\nu^{-1})$, and a wrapped Cauchy distribution $W^C\sim C(0,\sigma\nu^{-1})$. In the limit $\nu\to\infty$ the shot noise $\eta(t)$ becomes, respectively, Gaussian 
$
	\int_0^\tau \eta^{G}(t)dt \sim \sqrt{2D\tau} N(0,1)
$ 
or Cauchy white noise
$
	\int_0^\tau \eta^{C}(t)dt \sim \sigma \tau C(0,1)
$.
Another interesting case is that of random phase resetting with a uniform distribution $W^U=(2\pi)^{-1}$. 
Equation \eqref{Eq:Continuity} can be rewritten in terms of the Fourier components of $F$ with $F(\vartheta)=\sum_{k=-\infty}^\infty f_k\exp[-ik\vartheta]$ and order parameters (circular moments) $z_k=\left\langle e^{ik\vartheta}\right\rangle_P$ and $w_k=\left\langle e^{ik\phi}\right\rangle_W$ as an infinite system of coupled ordinary differential equations 
\begin{equation}\label{Eq:MomentsODEs}
	\dot z_k = ik\sum_{l=-\infty}^\infty f_{-l} z_{k+l}+ \nu \left(w_{k}-1\right) z_k.
\end{equation}
The circular moments $w_k^G$ of the wrapped Gaussian distribution, $w_k^C$ of the wrapped Cauchy distribution and $w_k^U$ of the uniform distribution are
\begin{equation}
    w^G_k = \exp\left(-\frac{D}{\nu}k^2\right),~w^C_k = \exp\left(-\frac{\sigma}{\nu}|k|\right),~ w_k^U = \delta_{k0}.~~
\end{equation}
In the limit $\nu\to\infty$ we recover the Fourier representation of the second derivative $\nu \left(w^G_{k}-1\right) \to -Dk^2$ for Gaussian white noise, and of the fractional derivative $\nu \left(w^C_{k}-1\right) \to -\sigma|k|$ for Cauchy white noise~\cite{Toenjes_etal-13,Chechkin_etal-03}. 

As a first result we show how the type of noise
affects stability of the asynchronous system state with a uniform phase distribution for Kuramoto-Daido coupling \eqref{Eq:KuraDaidoMF} where the Fourier modes of the force $F$ depend on the moments of the phase distribution as $f_k = h_kz_k$. In this case the incoherent state $P(\vartheta)=(2\pi)^{-1}$ or $z_{k}=\delta_{k0}$ is always a stationary solution of \eqref{Eq:MomentsODEs}. Linearization yields decoupled equations for different order parameters
\begin{equation}
	\dot z_k =  \left[ik\left(h_{k} + f_0 \right) + \nu \left(w_{k}-1\right)\right]z_k.
\end{equation}
The condition for stability of a mode $k>0$ is that the real part of the factor on the right hand side is smaller than zero, i.e.
\begin{equation}\label{Eq:IncoherentLinStab}
	\textrm{Re}\left[ik h_{k}+\nu\left(w_{k}-1\right)\right]<0.
\end{equation}
This gives three different stability criteria: 
$
	\textrm{Im}\left[\bar{h}_k/k\right] < D
$ for Gaussian white noise, 
$
	\textrm{Im}\left[\bar{h}_k\right] < \sigma
$ for Cauchy white noise, 
and 
$
	\textrm{Im}\left[k \bar{h}_k\right]<\nu
$
for random phase resetting.
Gaussian white noise, which diffuses phases on continuous trajectories, is most efficient in suppressing higher wave number instabilities by a factor $1/k$, whereas such instabilities occur more likely for strongly anharmonic coupling functions and Cauchy noise or random phase resetting. Phase coupling functions of that kind, e.g. with discontinuities or 
dead zones~\cite{Ashwin-Bick-Poignard-19} are sometimes constructed to design synchronization behavior in ensembles of artificial oscillators. 

We now demonstrate the main findings of the paper - 
that under Cauchy white noise, finite-dimensional reductions for a generally infinite system of mode dynamics \eqref{Eq:MomentsODEs} is possible.  The simplest case is that of sinusoidal forcing. The paradigmatic Kuramoto-Sakaguchi model, among others, belongs to this class. 
In terms of the phase dynamics~\eqref{Eq:PhaseEquations} this means $F(\vartheta) = \omega + \varepsilon\sin(\alpha-\vartheta)$ with generally time dependent parameters $\omega$, $\varepsilon$ and $\alpha$. Then \eqref{Eq:MomentsODEs} reduces to
\begin{equation}\label{Eq:TriRecurrence}
	\dot z_k = ik \left(f_1 z_{k-1} +f_0 z_k + f_{-1} z_{k+1}\right) -\sigma|k|z_k.
\end{equation}
One can straightforwardly check that this infinite system admits
the Ott-Antonsen (OA) ansatz $z_k=Z^k$ for $k\geq 0$.
The dynamics of the mean field $Z=z_1$ on this 
manifold is the OA equation
\begin{equation}\label{Eq:OA_ODE}
	\dot Z = i \left(f_1 + f_0 Z + f_{-1} Z^2\right) - \sigma Z.
\end{equation}
We stress here, that in our formulation
Eq.\,\eqref{Eq:OA_ODE} appears for \textit{identical} oscillators subject to independent Cauchy noise, while in the original OA formulation the same equation has been derived for \textit{noiseless, nonidentical}
oscillators with a Cauchy (Lorentz) distribution of natural frequencies.  Parameter $\sigma$ in the latter case characterizes the width of the distribution~\cite{Ott08}, in our case this parameter characterizes the noise intensity. 
The OA manifold corresponding to the family of wrapped Cauchy distributions (WCDs)
\begin{equation}
P_{OA}(\vartheta)=\frac{1}{2\pi}\frac{1-|Z|^2}{|e^{i\vartheta}-Z|^2}
\label{eq:wcd}
\end{equation}
where the dynamics \eqref{Eq:OA_ODE} is exact, has been shown to be globally attractive under \eqref{Eq:TriRecurrence} for oscillators with Cauchy distributed random frequencies \cite{Ott09}. This is therefore also true for identical oscillators subject to independent Cauchy white noise. 

However, the equivalence between oscillator ensembles with Cauchy noise and Cauchy distribution of natural frequencies, is no longer valid if \textit{higher harmonics} are present in the forcing $F(\vartheta)$. Then the first order parameter 
$z_1$ is coupled to $z_{-1}=\bar{z}_1$ through the mode $f_2$ in \eqref{Eq:MomentsODEs}, and can therefore not be an analytic function of the oscillator frequency - an essential condition in the OA approach. The averaging of the $z_k=z_k(\omega)$ over the frequency distribution $g(\omega)$ via application of the Cauchy integral theorem, which results in Eq.\,\eqref{Eq:TriRecurrence} if only the first harmonics are present, is no longer possible.
In contradistinction, for Cauchy noise the terms $-|k|\sigma$  
appear in the ODEs without the necessity of averaging over a frequency distribution, i.e. the equations are indeed valid for any multi-mode coupling. 

Our next goal is to generalize the OA approach for white
Cauchy noise to the situation where the coupling term
$F(\vartheta)$ contains up to $L$ harmonics. Then Eqs.~\eqref{Eq:MomentsODEs} become
\begin{equation}\label{Eq:CauchyODEs}
	\dot z_k = ik\left(f_0z_k+\sum_{l=1}^L \left[f_{l} z_{k-l} + f_{-l} z_{k+l}\right]
	\right) 
	-\sigma|k| z_k.
\end{equation}
The theory below does not provide a low-dimensional reduction of the dynamics \eqref{Eq:CauchyODEs}, but yields a low-dimensional
description of possible stationary solutions.
For phase densities which are stationary in a rotating reference frame, i.e. $\st{P}(\vartheta,t)=\st{P}(\vartheta-\Omega t)$, we have $\st{\dot z}_k = ik\Omega \st{z}_k$ with a yet unknown frequency offset $\Omega$. Through a shift into a co-rotating reference frame we can absorb $f_0$ in the frequency offset $\Omega\to f_0+\Omega$. The recurrence relations \eqref{Eq:CauchyODEs} for the stationary solution $\st{z}_k$ with $k\ge 1$ become independent of $k$
\begin{equation}\label{Eq:CauchyRecurrence}
	0 = \sum_{l=1}^{L} \left[f_{l} \st{z}_{k-l} + f_{-l} \st{z}_{k+l}\right] + \left(i\sigma-\Omega\right)  \st{z}_k.
\end{equation}
Recurrence relations of that kind are solved via the transfer matrix method. They have the general solution
\begin{equation}\label{Eq:GeneralSolution}
	 \st{z}_k = \sum_{m=1}^{2L} c_m \lambda_m^k, \qquad k\ge 1-L
\end{equation}
where complex factors $\lambda_m$ are the $2L$ roots of the characteristic equation
\begin{equation}\label{Eq:SymmetricSelfConsistency}
	\sum_{l=1}^L \left[f_{l}\lambda^{-l} + f_{-l}\lambda^{l}\right] = \Omega-i\sigma.
\end{equation}
Strictly speaking ansatz \eqref{Eq:GeneralSolution} is only valid for roots $\lambda_m$ of multiplicity one, which is almost always true if the $f_l$ are fixed. Since $k\ge 1$ in \eqref{Eq:CauchyRecurrence}, the ansatz \eqref{Eq:GeneralSolution} extends up to negative Fourier modes $k\ge 1-L$. Using the symmetric version of Rouch\'e's theorem one can show that if $\sigma \ne 0$, the number of solutions within the unit circle does not depend on the right hand side. Furthermore, if $\lambda$ is a root of the Laurent polynomial on the left hand side, then $\bar{\lambda}^{-1}$ is also a root, i.e. there are $L$ solutions of \eqref{Eq:SymmetricSelfConsistency} inside and $L$ solutions outside of the unit circle. Since the order parameters $z_k$ are bounded $|z_k|<1$, only the $\lambda_m$ with $|\lambda_m|<1$ contribute to the sum \eqref{Eq:GeneralSolution}. 
Given the $L$ roots $\lambda_m$ with $|\lambda_m|<1$, the coefficients $c_m$ are the unique solutions of the following set of linear equations for $k=1\dots L-1$
\begin{equation}\label{Eq:ck_linear}
	\sum_{m=1}^L c_m = 1, \qquad \sum_{m=1}^L c_m \lambda_m^k - \bar{c}_m\bar{\lambda}_m^{-k}=0.
\end{equation}
The first inhomogeneous equation expresses $z_0=1$, and the $L-1$ homogeneous equations are due to the conditions $\st{z}_k = \st{\bar{z}}_{-k}$ expressed in terms of \eqref{Eq:GeneralSolution} for $k=1\dots L-1$. This means all $z_k$ are fully determined by the set of eigenvalues $\lambda_m$. This is the desired low-dimensional reduction: possible stationary distributions in a population of oscillators forced in $L$ harmonics are
fully determined by $L$ complex parameters $\lambda_m$ with $|\lambda_m|<1$. Furthermore, \eqref{Eq:SymmetricSelfConsistency} constitutes a linear set of equations for the Fourier modes $f_l$, the noise intensity $\sigma$ and the frequency $\Omega$ which are thus known explicitly as functions of the $\lambda_m$.

Next, we demonstrate that this low-dimensional solution for a stationary  phase density is in fact a product of WCDs, which can be dubbed poly-WCD (cf.~\cite{Bian-Dickey-96})
\begin{equation}\label{Eq:StationaryPhaseDensity}
	\st{P}(\vartheta)  = \frac{1}{M}\frac{1}{2\pi}\prod_{m=1}^L \frac{1-\left|\lambda_m\right|^2}{\left|e^{i\vartheta}-\lambda_m\right|^2}\;.
\end{equation}
Given the values of $\lambda_m$ we explicitly calculate the coefficients $c_m$
\begin{eqnarray}
	\st{z}_k &=&\frac{1}{M}\int_0^{2\pi} \frac{e^{ik\vartheta}}{2\pi}\prod_{m=1}^L \frac{1-\left|\lambda_m\right|^2}{\left|e^{i\vartheta}-\lambda_m\right|^2} d\vartheta \\
	&=& \frac{1}{2\pi i} \oint\limits_{|z|=1} z^{k}\left[\frac{1}{M}\prod_{m=1}^L \frac{z\left(1-\left|\lambda_m\right|^2\right)}{(z-\lambda_m)(1-\bar{\lambda}_mz)}\right]\frac{dz}{z} \nonumber \\
	&=& \sum_{m=1}^L  \lambda_m^k \frac{1}{M} \prod_{n\ne m}^L \frac{\lambda_m\left(1-|\lambda_n|^2\right)}{\left(\lambda_m-\lambda_n\right)\left(1-\bar{\lambda}_n\lambda_m\right)}\quad(k\ge 1-L).\nonumber
\end{eqnarray}
The integrand has exactly the $L$ poles $\lambda_m$ on the unit disc if $k\ge 1-L$.
This begets \eqref{Eq:GeneralSolution} with coefficients
\begin{equation}\label{Eq:SelfConsistency03}
	c_m = \frac{1}{M}\prod\limits_{n\ne m}^L \frac{\lambda_m\left(1-|\lambda_n|^2\right)}{\left(\lambda_m-\lambda_n\right)\left(1-\bar{\lambda}_n\lambda_m\right)}
\end{equation}
and with normalizing weight
\begin{equation}\label{Eq:SelfConsistency04}
	M=\sum\limits_{l=1}^L \prod\limits_{p\ne l}^L \frac{\lambda_l\left(1-|\lambda_p|^2\right)}{\left(\lambda_l-\lambda_p\right)\left(1-\bar{\lambda}_p\lambda_l\right)}.
\end{equation}
Equations \eqref{Eq:SelfConsistency03}-\eqref{Eq:SelfConsistency04} together with 
\begin{equation}\label{Eq:SelfConsistency02}
    \st{z}_k = \sum_{m=1}^L c_m \lambda_m^k,    \qquad k\ge 1-L
\end{equation}
and Eq.\,\eqref{Eq:SymmetricSelfConsistency}
(where only roots with modulus smaller than 1 are taken)
form the self consistency conditions for the stationary phase density of identical phase oscillators subject to Cauchy white noise and under a forcing $F(\vartheta)$ with $L$ harmonics.  Equation~\eqref{Eq:SelfConsistency02} can be regarded as a generalization of the Ott-Antonsen ansatz, although it is restricted to stationary solutions. While algebraic self-consistency equations still require numeric root finding, the evaluation of these equations is much faster and numerical errors are much smaller than for integral self-consistency equations. Before proceeding to an example we mention that a circle distribution having moment $z^k=c\lambda^k$ has been considered by Kato and Jones~\cite{Kato-Jones-15}. Expression \eqref{Eq:SelfConsistency02} means that our stationary distributions are 
weighted sums of Kato-Jones distributions.

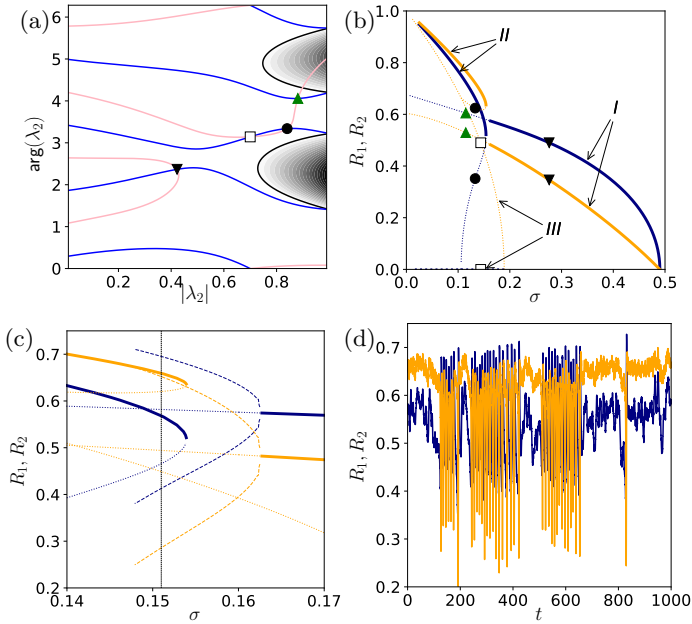
\begin{figure}[t]
\centering
\setlength{\unitlength}{1cm}
\begin{picture}(4.2,4.2)
\put(-0.0,0){\includegraphics[height=4.2cm]{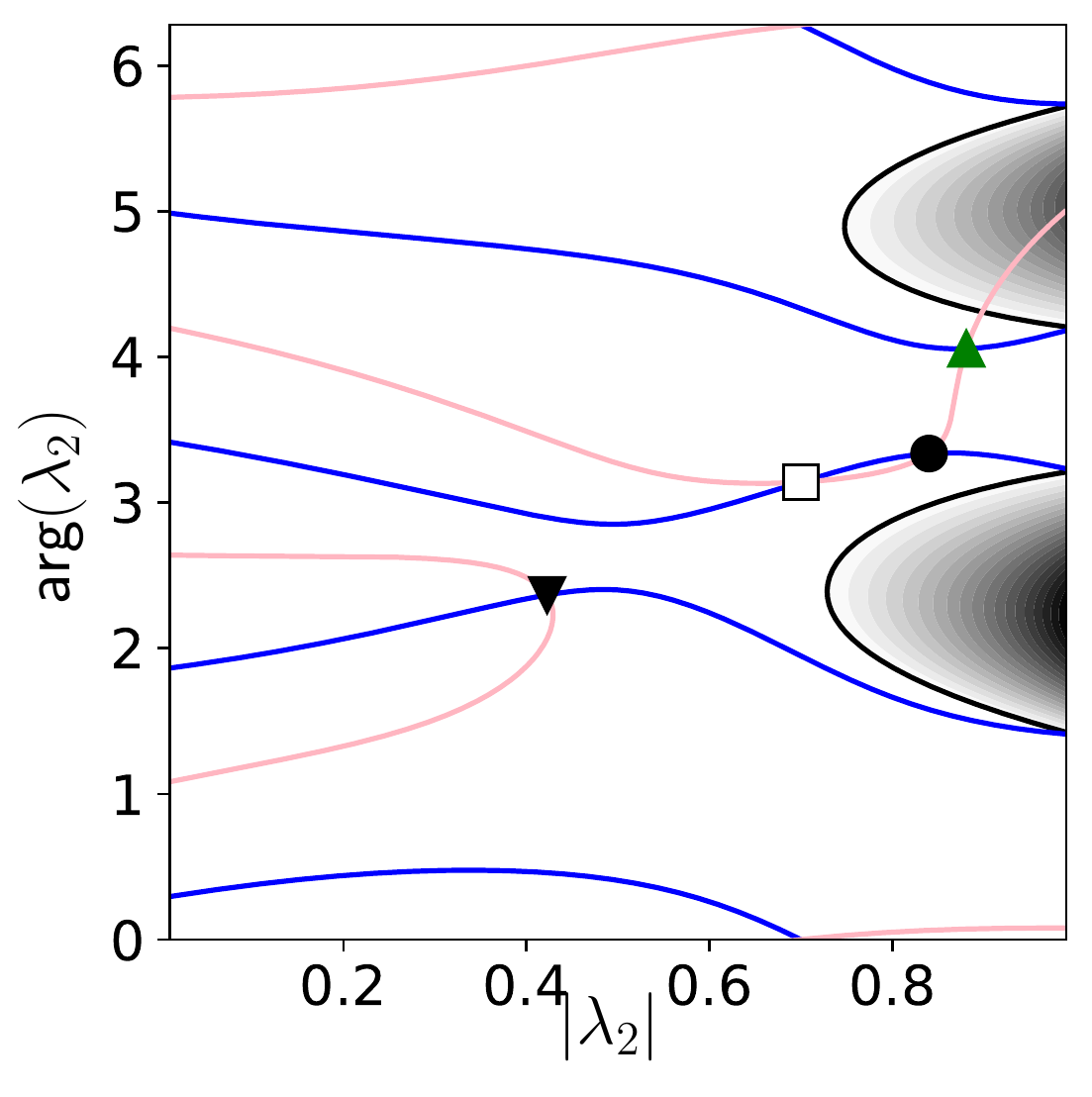}}
\put(0,3.8){(a)}
\end{picture}
\begin{picture}(4.2,4.2)
\put(-0.0,0){\includegraphics[height=4.2cm]{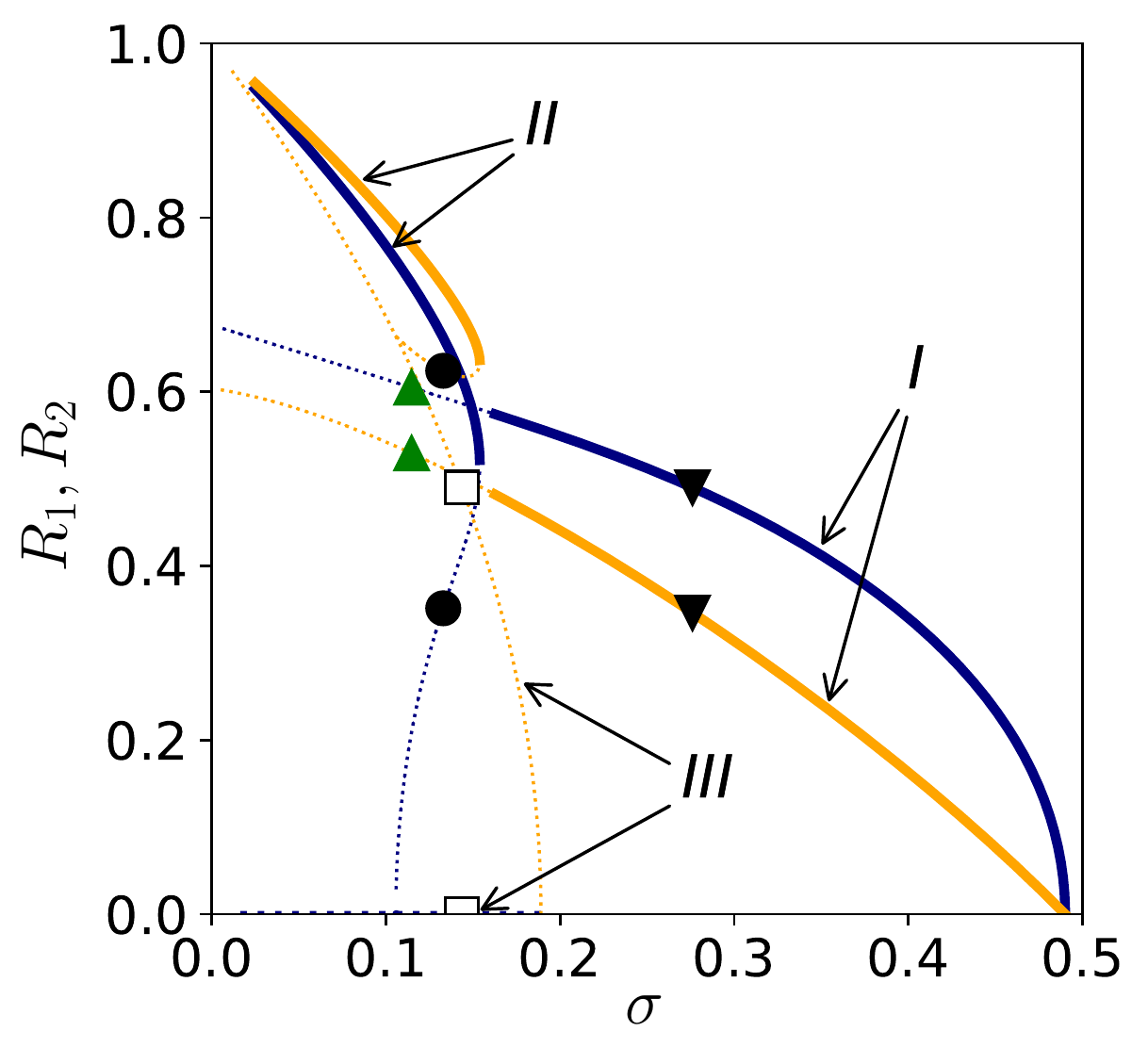}}
\put(0,3.8){(b)}
\end{picture}
\begin{picture}(4.2,4.2)
\put(-.2,0){\includegraphics[height=4.1cm]{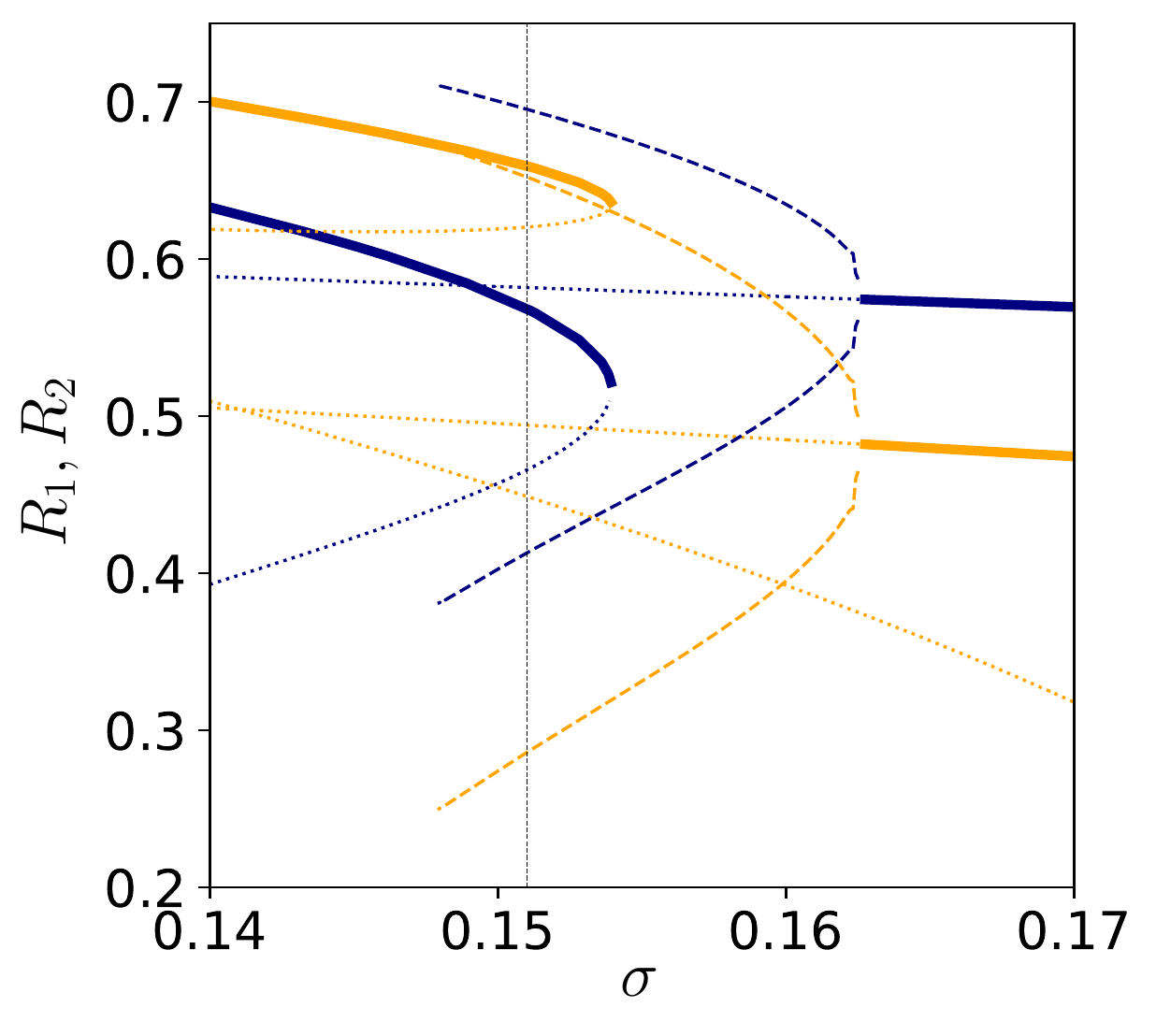}}
\put(-0.2,3.8){(c)}
\end{picture}
\begin{picture}(4.2,4.2)
\put(-0.0,0){\includegraphics[height=4.2cm]{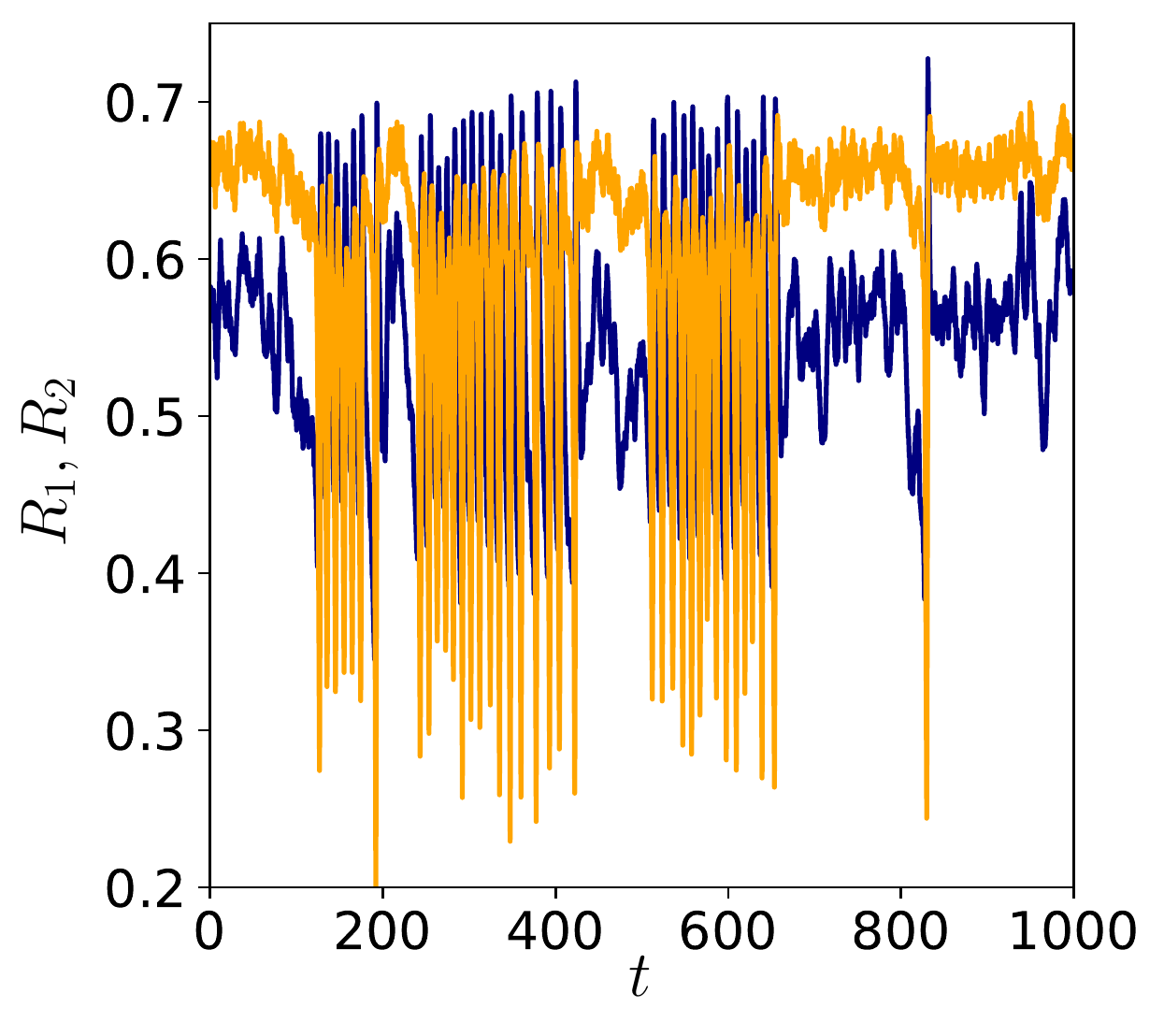}}
\put(0,3.8){(d)}
\end{picture}
\caption{\small (a) Zero lines for the imaginary (dark blue) and the real part (light red) of $\Delta$ 
as a function of $\lambda_2$ (amplitude and complex argument) given $\lambda_1=0.7$, $\varepsilon_1=1.0$, $\varepsilon_2=0.7$, $\alpha_1=-0.2$ and $\alpha_2=-1.0$. The points $\lambda_2\ne \lambda_1$ where these lines cross (marker symbols) correspond to stationary solutions. In the shaded region $\sigma<0$. (b) Collection of stationary solutions upon variation of $\lambda_1$ from zero to one. The dark blue and the light yellow lines mark the order parameters $R_1,R_2$, respectively. 
Bold lines are stable solutions and dotted lines unstable solutions. The markers correspond to the same stationary solutions depicted in (a), they 
are located on the three different branches (triangle markers are on the same branch).
(d) Zoom of the transition region from branch $I$ to branch $II$. Branch $I$ becomes unstable in a supercritical Hopf bifurcation when $\sigma\approx 0.163$. The dashed lines mark the maxima and the minima of the order parameters of the periodic solution when $\sigma$ is slowly decreased. Between $0.154<\sigma<0.163$ no stationary solution exists whereas between $0.148<\sigma<0.154$ the periodic solution co-exists with a stable equilibrium. Below $\sigma<0.148$ the periodic solution has disappeared in a SNIC bifurcation. (d) Integration of the Langevin equations \eqref{Eq:PhaseEquations} for $N=5000$ phase oscillators with phase difference coupling in the first an second harmonics and Cauchy noise of strength $\sigma=0.151$ in a small bistable regime. Stochastic switching between stationary and oscillating order parameters is observed.  
}\label{Fig1}
\end{figure}
Let us discuss the simplest nontrivial example: an ensemble of phase oscillators with a phase difference coupling \eqref{Eq:KuraDaidoMF} 
\begin{equation}
	H(\Delta\vartheta) = \varepsilon_1 \sin(\alpha_1-\Delta\vartheta) + \varepsilon_2\sin(\alpha_2-2\Delta\vartheta)
\end{equation}
in the first and the second harmonics, i.e. $h_k = \frac{1}{2i}\varepsilon_ke^{i\alpha_k} $, subject to Cauchy white noise. 
According to \eqref{Eq:SelfConsistency02}, stationary rotating wave solutions have the form
\begin{equation}
	\st{z}_k = c_1 \lambda_1^k + c_2 \lambda_2^k
\end{equation}
where, as it follows from \eqref{Eq:SelfConsistency03} 
\begin{equation}
	c_1 = 1-c_2 = \left[1-\frac{\lambda_2(1-|\lambda_1|^2)(1-\bar{\lambda}_2\lambda_1)}{	\lambda_1(1-|\lambda_2|^2)(1-\bar{\lambda}_1\lambda_2)}\right]^{-1}
\end{equation}
and from \eqref{Eq:SymmetricSelfConsistency}, $\lambda_1$ and $\lambda_2$ are simultaneous solutions of the algebraic equations
\begin{eqnarray}\label{Eq:SymmetricSelfConsistency_L2}
	\Omega-i\sigma &=& 
	\bar{h}_2\st{\bar{z}}_2\lambda_{1,2}^{2} +
	\bar{h}_1\st{\bar{z}}_1\lambda_{1,2}+
	h_1\st{z}_1\lambda_{1,2}^{-1} +
	h_2\st{z}_2\lambda_{1,2}^{-2}.\qquad
\end{eqnarray}
One parameter in the problem can be eliminated by a rescaling of 
time. In this example we choose $\varepsilon_1=1$. 
Furthermore, because of rotational invariance $\vartheta\to\vartheta+const$, we can choose $\lambda_1\in[0,1]$ to be real. Given this free parameter we can calculate the right hand sides of~\eqref{Eq:SymmetricSelfConsistency_L2}
as functions of $\lambda_2$. At points $\lambda_2\ne\lambda_1$ where the difference $\Delta$ between the r.h.s. of \eqref{Eq:SymmetricSelfConsistency_L2} vanishes, a stationary solution exists in a rotating reference frame with frequency $\Omega$ and noise $\sigma$, which are thus defined parametrically as the real part and the negative imaginary part of the r.h.s. of \eqref{Eq:SymmetricSelfConsistency_L2}, respectively. Tuning $\lambda_1$ from zero to one we can quickly pinpoint all $\lambda_2$ where the difference $\Delta$ is zero (see Fig.\,\ref{Fig1}(a)), and continue these different solution branches. We illustrate the found bifurcation diagram in Fig.\,\ref{Fig1}. Zeros of $\Delta$ found according to Fig.\,\ref{Fig1}(a) result in three branches of solutions. Branch $I$, starting at the point of instability of the first mode $\sigma_{c1}=\cos(0.2)/2\approx 0.49$,
is a state with a bi-modal distribution, where both real order parameters $R_{1,2}=|z_{1,2}|$
are non-zero. Another branch $III$, which starts at the instability point of the second mode $\sigma_{c2}=0.7\cos(1)/2\approx 0.19$, is a pure symmetric bi-modal
solution, where only even modes are present and $R_1=0$. This branch in terms of the eigenvalues $\lambda_{1,2}$ correspond to the case
$\lambda_1=-\lambda_2$, $c_1=c_2=\frac{1}{2}$. There is also a third nontrivial mode $II$ that bifurcates at $\sigma_{c3}\approx 0.1$ from the incoherent branch $III$ and folds back at larger values of $\sigma$ to become the unique stationary solution in the limit of small noise strength.
Stability of stationary states (and periodic solutions) was checked numerically from \eqref{Eq:CauchyODEs} with truncation at a large number of Fourier modes.
Remarkably, there is a stability change from mode $I$ to mode $II$ (mode $III$ is always unstable). 

\begin{figure}[t]
\centering
\setlength{\unitlength}{1cm}
\begin{picture}(8.4,3.6)
\put(0,-0.4){\includegraphics[height=3.9cm]{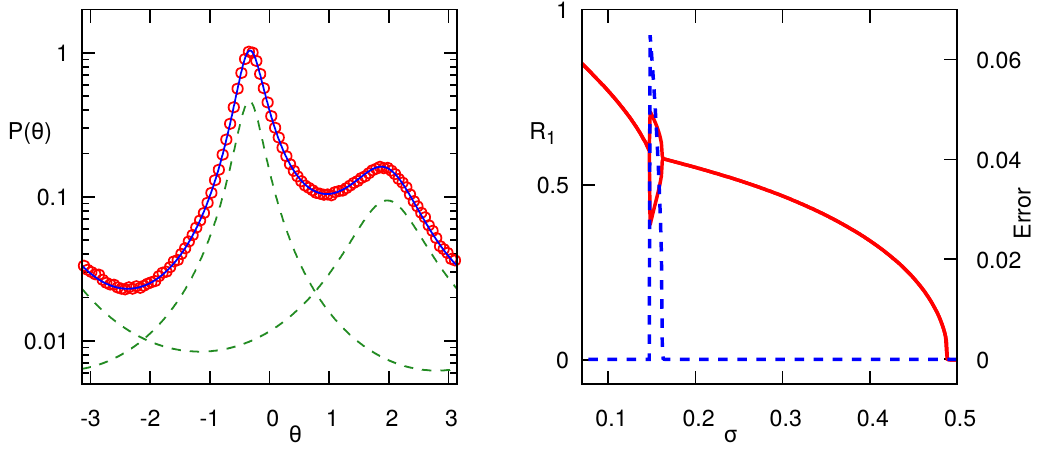}}
\put(0,3.2){(a)}
\put(4.2,3.2){(b)}
\end{picture}
\caption{(a) Comparison of the empirical probability density obtained
in simulations of an ensemble of $N=10^4$ oscillators (red circles)
with theoretical prediction of poly-WCD (solid curve under the circles), for branch $I$ 1 at $\sigma=0.2$. Dashed green lines show
corresponding wrapped Cauchy densities (arbitrarily scaled).
(b) Bifurcation diagram (red: span between $R_{1max}$ and $R_{1min}$) obtained from the solutions of system~\eqref{Eq:CauchyODEs}
at slowly decreasing parameter $\sigma$. Dashed blue line (see scale at the right) shows the deviation from the poly-WCD.}
\label{fig:td}
\end{figure}

An interesting feature of the bifurcation diagram Fig.\,\ref{Fig1}
is that there is a range of noise intensities $\sigma$ where all steady states
are unstable, and stable periodic oscillations of the order parameters are observed. Furthermore, there is an even smaller range of bi-stability where stable periodic oscillations coexists with the stable stationary state of  branch $II$. Only for stationary states we do expect validity of the poly-WCD distribution. To characterize this, it is instructive to consider a Fourier transform of the logarithm of $\st{P}(\vartheta)$ \eqref{Eq:StationaryPhaseDensity}. In our example this is $\ln\st{P}(\theta)=\text{const}+\sum_{k>0} (s_ke^{-ik\vartheta}+\bar{s}_k e^{ik\vartheta})$ with $s_k=k^{-1}(\lambda_1^k+\lambda_2^k)$. Remarkably, this representation does not contain the constants $c_m$. Because all Fourier modes $s_k$ depend 
on two complex numbers only, Fourier modes 
with $k>2$ can be represented through $s_{1,2}$. In particular, $s_3=s_1s_2-s_1^3/6$. Thus, quantity $\text{Err}=|s_3-s_1s_2+s_1^3/6|$ serves as a measure
of the deviation from the poly-WCD. We show this deviation together with the empirical bifurcation diagram in Fig.\,\ref{fig:td}(b).
One can see that indeed, while stationary states are given by \eqref{Eq:StationaryPhaseDensity}, a clear deviation occurs for periodic solutions.

In conclusion, we have demonstrated that an ensemble of identical phase oscillators subject to independent Cauchy white noise admits a finite-dimensional description. In the simplest case of sinusoidal forcing, the resulting reduction is just the Ott-Antonsen ansatz with a wrapped Cauchy distribution of the phases. If forcing contains up to $L$ Fourier modes, stationary states are given by a poly-wrapped Cauchy distribution with $L$ complex roots $\lambda_m$ on the unit disc as parameters. This finite-dimensional reduction is valid not only for Kuramoto-Daido type coupling, but also for more general situations, such as Winfree-type models and ensembles with nonlinear coupling. 

We stress here that the special role of the Cauchy distribution is well-known in statistical physics, starting from the seminal work by Lloyd on the exact solution~\cite{Lloyd_1969}
for disordered Hamiltonians with Cauchy distributed heterogeneities. In the context of populations of dynamical elements, this property has been explored in studies of homographic maps~\cite{Griniasty-Hakim-94}.
Our study shows that the Cauchy noise significantly simplifies the description of the dynamics, compared to the Gaussian noise, also for populations of phase oscillators.
\begin{acknowledgments}
We thank D. Goldobin for fruitful discussions. A. P.
acknowledges support by the Russian Science Foundation (grant Nr. 17-12-01534) and by DFG (grant PI 220/21-1).
\end{acknowledgments}
\bibliography{biblio}
\end{document}